\documentclass[sigconf,screen,9pt]{acmart}
\AtBeginDocument{%
  }

\copyrightyear{2026}
\acmYear{2026}
\setcopyright{cc}
\setcctype{by}
\acmConference[DAC '26]{63rd ACM/IEEE Design Automation Conference}{July 26--29, 2026}{Long Beach, CA, USA}
\acmBooktitle{63rd ACM/IEEE Design Automation Conference (DAC '26), July 26--29, 2026, Long Beach, CA, USA}
\acmDOI{10.1145/3770743.3804146}
\acmISBN{979-8-4007-2254-7/2026/07}
\begin{document}

\title{ChatSVA: Bridging SVA Generation for Hardware Verification via Task-Specific LLMs}

\author{%
\parbox{\textwidth}{\centering
Lik Tung Fu\textsuperscript{1,2},
Jie Zhou\textsuperscript{1,2},
Shaokai Ren\textsuperscript{2},
Mengli Zhang\textsuperscript{2},
Jia Xiong\textsuperscript{1,2},\\[2pt]
Hugo Jiang\textsuperscript{1,2},
Nan Guan\textsuperscript{3},
Xi Wang*\textsuperscript{1,2},
Jun Yang\textsuperscript{1,2}
}}

\affiliation{
  \institution{
    \small
    \textsuperscript{1}National ASIC Center, School of Integrated Circuits, Southeast University, China \\
    \textsuperscript{2}National Center of Technology Innovation for Electronic Design Automation, China \\
    \textsuperscript{3}Department of Computer Science, City University of Hong Kong, Hong Kong, China \\
    Email: liktungfu@seu.edu.cn, 230240053@seu.edu.cn, renshaokai@nctieda.com, zhangmengli@nctieda.com, xiongjia@seu.edu.cn, 101013615@seu.edu.cn, 
    nanguan@cityu.edu.hk, xi.wang@seu.edu.cn, dragon@seu.edu.cn \\
    \textsuperscript{*}Corresponding author.
  }
  \country{}
}

\renewcommand{\shortauthors}{Fu et al.}

\begin{abstract}
% Functional verification accounts for over 50\% of the IC development lifecycle, making SystemVerilog Assertions (SVAs) indispensable for rigorous digital chip verification. SVAs enable comprehensive formal property checking and significantly enhance simulation-based debugging. 
Functional verification consumes over 50\% of the IC development lifecycle, where SystemVerilog Assertions (SVAs) are indispensable for formal property verification and enhanced simulation-based debugging.
However, manual SVA authoring is labor-intensive and error-prone. 
% While Large Language Models (LLMs) offer potential for automating verification tasks, their direct deployment is hindered by low functional accuracy and severe scarcity of domain-specific SVA data, limiting their effectiveness.
While Large Language Models (LLMs) show promise, their direct deployment is hindered by low functional accuracy and a severe scarcity of domain-specific data.
To address these challenges, we introduce ChatSVA, an end-to-end SVA generation system built upon a multi-agent framework. At its core, the AgentBridge platform enables this multi-agent approach by systematically generating high-purity datasets, overcoming the data scarcity inherent to few-shot scenarios. 
% Evaluated on 24 RTL designs, ChatSVA achieves the syntax and functional pass rates of \textbf{98.66\%} and \textbf{96.12\%}, respectively, surpassing privous state-of-the-art (SOTA) with a improvement of 33.3\% in functional correctness. On average, it generates \textbf{139.5 SVAs per design} with an \textbf{82.50\%} Function Coverage, translating to over a \textbf{11$\times$} enhancement in bug detection capability compared to the privous SOTA.
Evaluated on 24 RTL designs, ChatSVA achieves \textbf{98.66\%} syntax and \textbf{96.12\%} functional pass rates, generating \textbf{139.5} SVAs per design with \textbf{82.50\%} function coverage. This represents a \textbf{33.3 percentage point} improvement in functional correctness and an over \textbf{11$\times$} enhancement in function coverage compared to the previous state-of-the-art (SOTA).
ChatSVA not only sets a new SOTA in automated SVA generation but also establishes a robust framework for solving long-chain reasoning problems in few-shot, domain-specific scenarios.
% A publicly accessible ChatSVA web service is available to the broader community and a demo is available for reviewers to comply with the double-blind review policy: https://anonymous.4open.science/r/ChatSVA-3502.
% An online service has been publicly released at \url{https://www.nctieda.com/CHATDV.html}.
A publicly accessible ChatSVA web service is available at \url{https://www.nctieda.com/CHATDV.html}.

\end{abstract}





\maketitle

%----------------------------------------------------
%-----------Introduction
%----------------------------------------------------
\section{Introduction}
\label {sec:introduction}

\begin{figure}[!t] % 浮动体定位控制符
  \centering
  \includegraphics[width=0.93\columnwidth]{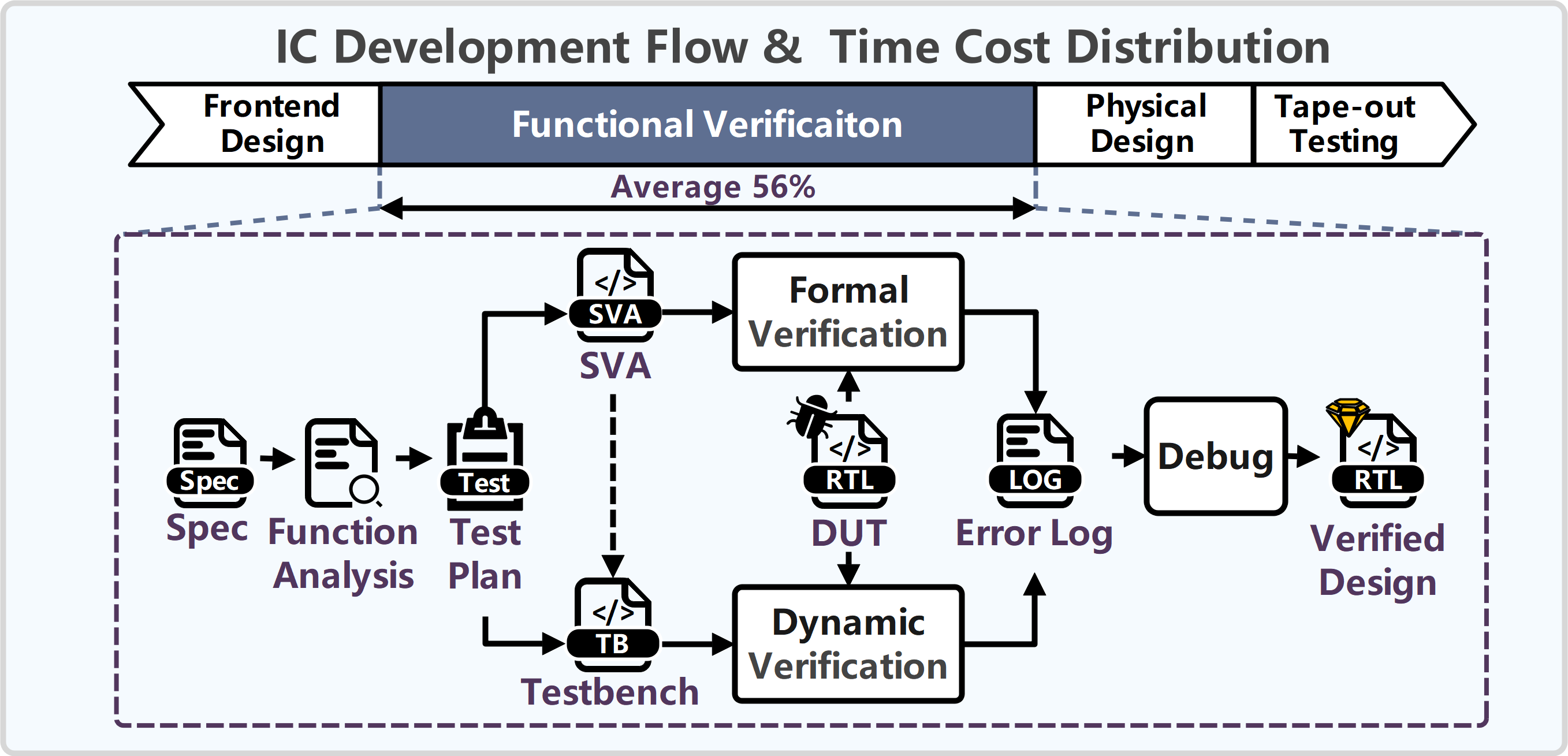} % 宽度适配单栏
  \caption[Short caption]{\small IC Development Flow \&  Time Cost Distribution. Functional verification accounts for 56\% of the development time~\cite{simens_report_2021}.}
  \label{fig:verification}
    \vspace{-2pt}
\end{figure}

\begin{figure*}[t]
  \centering
  \includegraphics[width=0.95\linewidth]{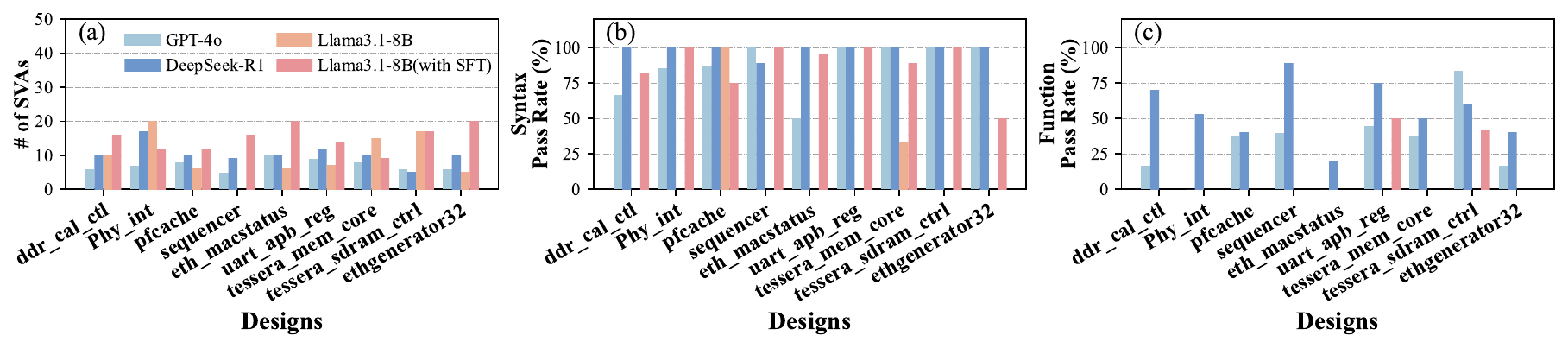}
  \caption{\small SVA Generation Capabilities of LLMs. (a) Number of SVAs.(b) Syntax Pass Rate.(c) Function Pass Rate.}
  \label{fig:bg}
    \vspace{-2pt}
\end{figure*}

The exponential increase in integrated circuit (IC) complexity, driven by domain-specific architectures and heterogeneous computing~\cite{simba,simens_report_2023,simens_report_2021,simens_report_2019}, has created a critical bottleneck: functional verification. Consuming over 50\% of the IC development lifecycle (Fig.~\ref{fig:verification}), this phase is the most labor-intensive stage in hardware design~\cite{Falsafi2023AgileVerification,DeliveringFunctionalVerification}. Traditional simulation is increasingly insufficient, leaving a high risk of costly post-silicon bugs~\cite{challenge,are_we_there}.

To address this, Assertion-Based Verification (ABV) using SystemVerilog Assertions (SVA) emerged as a standard practice~\cite{hasan2015formal,revolution}. SVAs serve as formal properties for exhaustive checking and as runtime monitors in simulation. Despite their power, manual SVA authoring has become a bottleneck, as the process is labor-intensive, error-prone, and requires expertise~\cite{surveyassertion}. Consequently, automating SVA generation is critical to break the verification deadlock.

The pursuit of SVA automation has seen various approaches. Early traditional natural language processing (NLP) attempts were hampered by shallow semantic understanding, unable to grasp complex design intent~\cite{GLAsT-NLP,subtreeNLP}. The emergence of Large Language Models (LLMs) marked a paradigm shift~\cite{chatcpu,Chip-chat,ai_assertion}, sparking new research that demonstrated a significant leap in generalization over predecessors~\cite{assertllm,AssertionForge}. However, a definitive breakthrough remains elusive, as current methods treat SVA generation as a single-step translation, ignoring its nature as a complex long-chain reasoning process. This flawed monolithic approach is compounded by a chronic scarcity of high-quality, domain-specific training data.

% The pursuit of SVA automation has seen various approaches. Early traditional natural language processing (NLP) attempts were hampered by shallow semantic understanding, unable to grasp complex design intent~\cite{GLAsT-NLP,subtreeNLP}. The emergence of Large Language Models (LLMs) marked a paradigm shift~\cite{chatcpu,Chip-chat,chattest,vgv,genben}, sparking new research that demonstrated a significant leap in generalization over predecessors~\cite{assertllm,AssertionForge}. However, a definitive breakthrough remains elusive, as current methods treat SVA generation as a single-step translation, ignoring its nature as a complex long-chain reasoning process. This flawed monolithic approach is compounded by a chronic scarcity of high-quality, domain-specific training data.

% Our contributions can be summarized as follows:
% Therefore, we introduce the ChatSVA framework to address these challenges with a two-part strategy. Our main contributions are:

% To address the aforementioned challenges, we introduce two core components that form our solution: ChatSVA, a multi-agent SVA generation framework, and AgentBridge, a principled data synthesis platform. Together, they enable a robust methodology for high-quality SVA generation. Our main contributions are:

To address these challenges, we present ChatSVA, a multi-agent SVA generation framework, together with AgentBridge, a data synthesis and augmentation platform. Our solution significantly enhances SVA generation in terms of Syntax Pass Rate, Function Pass Rate, and Function Coverage. Our main contributions are:
\begin{itemize}
    \item We introduce \textbf{ChatSVA}, an end-to-end, multi-agent framework that decomposes the long-chain reasoning of SVA generation, significantly enhancing functional correctness and coverage.

    \item We introduce \textbf{AgentBridge}, a data synthesis platform designed to generate high-purity, verifiable dataset, effectively solving the data scarcity problem in few-shot scenarios.
 
    \item We employ a training strategy combining Supervised Fine-Tuning (SFT) and Retrieval-Augmented Generation (RAG), demonstrating state-of-the-art (SOTA) generative capabilities on a comprehensive benchmark.
    
    \item We have conducted extensive experiments to validate the effectiveness of our framework and have released an online service for public access and reproducibility.
\end{itemize}

\section{Motivation \& Related Work}
% \section{Motivation}
\label{sec:background}

\subsection{Traditional SVA Design}
Automating SVA generation has been a long-standing challenge. Early efforts followed two main paths. Dynamic methods ``mine'' SVAs from simulation traces~\cite{harm,ateam,goldmine}, but were fundamentally flawed, depending on buggy DUTs and requiring testbench stimulus, thus forfeiting formal verification benefits. Static methods, on the other hand, used traditional NLP to parse the specification (Spec). Despite numerous attempts to improve their generalization capabilities with enhanced frameworks~\cite{GLAsT-NLP,subtreeNLP,CNLNLP,AutomatedNLP} or hybrid machine learning techniques~\cite{ChatbotNLP,HybridNLP,SpectosvaNLP}, they were consistently hampered by the limited semantic understanding of pre-LLM methods. 

% Automating SVA generation has been a long-standing challenge. Early efforts followed two main paths. Dynamic methods ``mine'' SVAs from simulation traces~\cite{harm,ateam,goldmine}, but were fundamentally flawed, depending on buggy Design Under Tests (DUTs) and requiring testbench stimulus, thus forfeiting formal verification benefits. Static methods, on the other hand, used traditional NLP to parse the specification (Spec). Despite numerous attempts to improve their generalization capabilities with enhanced frameworks~\cite{GLAsT-NLP,subtreeNLP,CNLNLP,AutomatedNLP} or hybrid machine learning techniques~\cite{ChatbotNLP,HybridNLP,SpectosvaNLP}, they were consistently hampered by the limited semantic understanding of pre-LLM methods. 

\subsection{LLM-aided SVA Generation}

The advent of LLMs marked a paradigm shift, offering a promising solution to the semantic gap that plagued earlier methods~\cite{meic,location,uvllm,chatchisel}. Current research has rapidly converged on two main strategies: (1) augmenting general-purpose models with prompt engineering or RAG~\cite{assertllm,AssertionForge,chiraag,Usingllm,securityllm}, and (2) fine-tuning smaller models with domain-specific data~\cite{validatablellm,finetunellm,towardllm,domainllm}. 

Despite significant improvements in syntactic correctness, the practical utility of these methods remains limited, primarily due to their reliance on a flawed, monolithic methodology. Most approaches treat SVA generation as a single-step translation from Spec to code~\cite{chiraag,Usingllm,validatablellm,finetunellm,towardllm}. This oversimplified approach forces a focus on ``syntactic correctness'' over ``functional efficacy'', as it lacks the intermediate steps needed to reason about complex design intent. Consequently, the generated SVAs are often functionally simplistic or incorrect, providing little meaningful verification coverage.
Furthermore, this methodological weakness is compounded by a risky practice: using the RTL code itself as an input to guide generation~\cite{AssertionForge,chiraag,securityllm}. This introduces ``verification contamination'', where the model may learn and replicate existing design flaws, undermining the very purpose of verification. Thus, despite recent progress, a reliable and safe methodology for LLM-aided SVA generation has yet to be established.

\begin{figure*}[!t]
  \centering
  \includegraphics[width=0.93\linewidth]{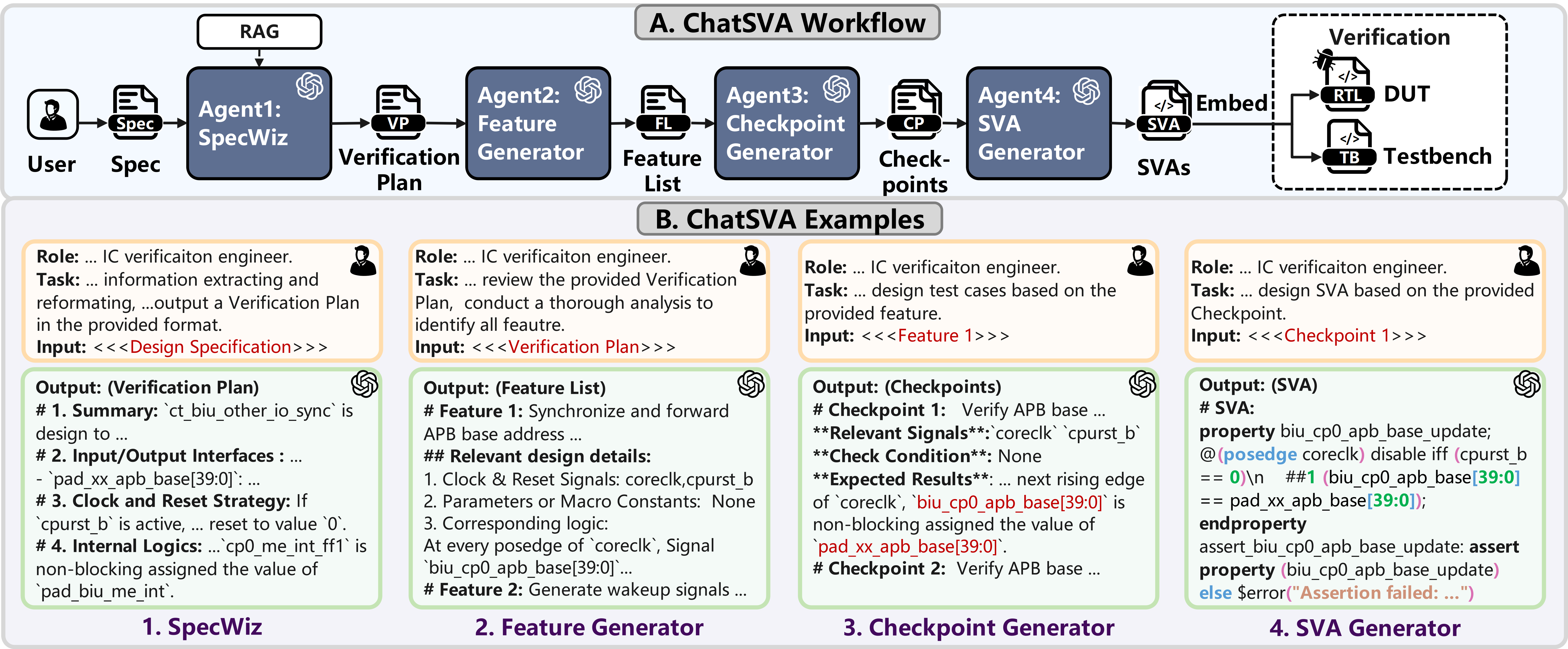}
  \caption{\small ChatSVA Workflow \& Examples}
  \label{fig:workflow}
    % \vspace{-2pt}
\end{figure*}

%The advent of LLMs marked a paradigm shift, offering a path to overcome prior limitations~\cite{chatcpu,Chip-chat}. Current research follows two main strategies: augmenting general-purpose models (e.g., with RAG)~\cite{assertllm,chiraag,Usingllm}, and fine-tuning domain-specific ones~\cite{validatablellm,finetunellm,towardllm}. While these approaches improved syntactic correctness, their practical utility remains limited. Their functional correctness is often insufficient for sign-off, and some methods even introduce new risks, such as using unverified RTL code as input, which can inject design flaws into the verification logic~\cite{AssertionForge}.

\subsection{Challenges}
To empirically test our critique of the prevailing monolithic strategy, we analyzed representative models (GPT-4o, DeepSeek-R1, and Llama3.1-8B base/SFT) on the FIXME benchmark~\cite{FIXME}. The experiment, designed to isolate the value of methodology from final data, yielded stark results (Fig.~\ref{fig:bg}): 
(1) The number of SVAs was low, never exceeding 20 SVAs per design. 
(2) While general-purpose models often achieved a high Syntax Pass Rate, sometimes 100\%, the fine-tuned model, despite significant gains over its base version, could not match them. 
(3) Most critically, the Function Pass Rate was poor across all models. 
These results reveal three fundamental challenges:

% \subsection{Challenges}
% To empirically test our critique of the prevailing monolithic strategy, we analyzed representative models (GPT-4o, DeepSeek-R1, and Llama3.1-8B w/SFT) on the FIXME benchmark~\cite{FIXME}. In particular, the SFT model was trained on the same final dataset used in our full system, ensuring that any remaining gap cannot be attributed to a lack of fine-tuning. The results are clear (Fig.~\ref{fig:bg}): 
% (1) The number of SVAs was low, never exceeding 20 SVAs per design. 
% (2) While general-purpose models often achieved a high Syntax Pass Rate, sometimes 100\%, the fine-tuned model, despite significant gains over its base version, could not match them. 
% (3) Most critically, the Function Pass Rate was poor across all models. 
% These results reveal three fundamental challenges:

% To test our critique of the prevailing monolithic strategy, we analyzed representative models (GPT-4o, DeepSeek-R1, and Llama3.1-8B base/SFT) on the FIXME benchmark~\cite{FIXME}. The experiment was designed to isolate the value of methodology from final data: we applied SOTA general models (GPT-4o, DeepSeek-R1) in a single step, while crucially fine-tuning model on the exact same final Spec-to-SVA pairs using in our work. The results, shown in Fig.~\ref{fig:bg}, revealed three fundamental challenges:

\noindent\textbf{C1: Inadequate Generation Capability.}
These results highlight a clear performance ceiling. The high Syntax Pass Rate of general-purpose models, contrasted with the universally low Function Pass Rate, reveals a critical disconnect: models achieve syntactic correctness but fail functionally. The limited improvement of the fine-tuned model proves that fine-tuning on the final output is ineffective at capturing deep functional logic. This functional failure, combined with low generation volume, confirms the monolithic approach severely limits practical generation capability.

% \noindent\textbf{C1: Inadequate Generation Capability.}
% These results highlight a clear performance ceiling. The high Syntax Pass Rate of general-purpose models, contrasted with the universally low Function Pass Rate, reveals a critical disconnect: models achieve syntactic correctness but fail functionally. The limited improvement of the fine-tuned model proves that fine-tuning on the final output is ineffective at capturing deep functional logic. This functional failure, combined with low generation volume, confirms the monolithic approach severely limits practical generation capability.

% The results highlight a clear performance ceiling for the monolithic approach. While the Llama3.1-SFT model demonstrated a significant performance boost over its base version, confirming the value of fine-tuning, its functional correctness remained far below that of the general-purpose models like GPT-4o. Furthermore, all models, including the top-tier ones, produced a starkly low volume of SVAs and exhibited brittle syntactic correctness. This proves that while fine-tuning is beneficial, it cannot overcome the inherent limitations of the single-step methodology.

\noindent\textbf{C2: Flawed Monolithic Methodology.}
C1 points to a methodological flaw: treating SVA generation as a monolithic Spec-to-SVA translation. This single-step approach forces models to prioritize syntactic correctness over functional efficacy, as it lacks the intermediate steps to reason about complex design intent. This forces a single, unguided inferential leap, making LLMs prone to ``functional hallucinations'', generating SVAs that are syntactically valid but functionally incorrect.

% \noindent\textbf{C2: Flawed Monolithic Methodology.}
% C1 points to a methodological flaw: treating SVA generation as a monolithic Spec-to-SVA translation. This single-step approach forces models to prioritize syntactic correctness over functional efficacy, as it lacks the task-decomposition steps to reason about complex design intent. This forces a single, unguided inferential leap, making LLMs prone to ``functional hallucinations'', generating SVAs that are syntactically valid but functionally incorrect.

%\noindent\textbf{C2: Flawed Monolithic Methodology.}
% This uniform failure points to a deeper methodological flaw: treating SVA generation as a monolithic Spec-to-SVA translation. In reality, it is a long-chain reasoning process requiring decomposition. By forcing a single, unguided inferential leap, LLMs are prone to functional hallucinations.

\noindent\textbf{C3: The Few-Shot Data Dilemma.} Hardware verification is inherently a few-shot problem due to scarce public data. This scarcity pushes approaches towards a monolithic methodology, as even collecting Spec-to-SVA pairs is challenging, let alone intermediate data. This creates a vicious cycle: the monolithic approach ignores the intermediate data needed to teach reasoning, trapping models into learning superficial mappings instead of deep verification logic.

% \noindent\textbf{C3: The Few-Shot Data Dilemma.} Hardware verification is inherently a few-shot problem due to scarce open-source data. This scarcity pushes approaches towards a monolithic methodology, as even collecting Spec-to-SVA pairs is challenging, let alone task-decomposition data. This creates a vicious cycle: the monolithic approach ignores the task-decomposition data needed to teach reasoning, trapping models into learning superficial mappings instead of deep verification logic.

% \noindent\textbf{C3: The Few-Shot Learning Data Dilemma.} The hardware verification domain is inherently a few-shot learning problem due to the scarcity of public, high-quality verification data. This data scarcity forces many approaches into a simplistic monolithic methodology, as collecting paired Spec-to-SVA data is already challenging, let alone the intermediate reasoning steps. This creates a vicious cycle: the monolithic approach ignores the crucial intermediate data required to teach a model \emph{how} to reason, trapping the field in a state where models can only learn superficial mappings, not deep verification logic.

% These challenges clarify why applying ever-larger LLMs yields diminishing returns and necessitate the fundamental shift in methodology that we propose.

%----------------------------------------------------
%-----------Design
%----------------------------------------------------

\section{Methodology}
\label{sec:Design}
The three challenges identified previously necessitate a fundamental shift in approach. To address these interconnected issues, we propose a two-pronged strategy designed to break the vicious cycle of methodological oversimplification and data scarcity.

\paragraph{\textbf{S1: Long-Chain Reasoning Decomposition.}}

% \noindent\textbf{S1: Long-Chain Reasoning Decomposition.}

S1 directly targets C2 and C1 by decomposing the monolithic task into a pipeline of modular sub-tasks. This provides the intermediate reasoning steps C2 identifies as missing, shifting the focus from syntactic correctness to functional intent. By replacing the single, unreliable leap with a series of verifiable steps, this strategy directly addresses the root cause of the low Function Pass Rate observed in C1. This is instantiated in our \textbf{ChatSVA} framework (Section~\ref{sec:ChatSVA}).

% S1 directly targets C2 and C1 by decomposing the monolithic task into a pipeline of modular sub-tasks. This provides the intermediate reasoning steps C2 identifies as missing, shifting the focus from syntactic correctness to functional intent. By replacing the single, unreliable leap with a series of verifiable steps, this strategy directly addresses the root cause of the low Function Pass Rate observed in C1. This is instantiated in our \textbf{ChatSVA} framework (Section~\ref{sec:ChatSVA}).

\paragraph{\textbf{S2: Data Synthesis for Few-shot Learning.}}

% \noindent\textbf{S2: Data Synthesis for Few-shot Learning.}

While S1 provides a principled decomposition of the design tasks, decomposition alone cannot overcome the data scarcity challenge identified in C3. To address this, we introduce \textbf{AgentBridge} (Section~\ref{sec:AgentBridge}), a unified data synthesis framework that generates high-quality, task-specific datasets for each agent in the ChatSVA pipeline.
Notably, AgentBridge provides a practical path to solving C1, where high functional correctness depends primarily on abundant and consistent training data.

\paragraph{\textbf{Synergy.}}

% \noindent\textbf{Synergy.} 
%S1 and S2 work in concert to break the vicious cycle of C3: 
% S1 provides the structural blueprint (\emph{what} to do), while S2 makes it executable with the necessary data (\emph{how} to do it), finally enabling a shift from syntactic plausibility to functional correctness in SVA generation.
Taken together, S1 and S2 form a complementary pair. S1 decomposes the overall task into structurally coherent units, clarifying \emph{what} needs to be generated at each stage. S2 then supplies the data required for each unit, enabling the agents to reliably execute \emph{how} it should be generated. This interplay between structural guidance and data support allows the pipeline to move beyond merely producing syntactically plausible outputs toward achieving functional correctness in SVA generation.
\begin{figure*}[h!] 

  \centering
  \includegraphics[width=0.92\linewidth]{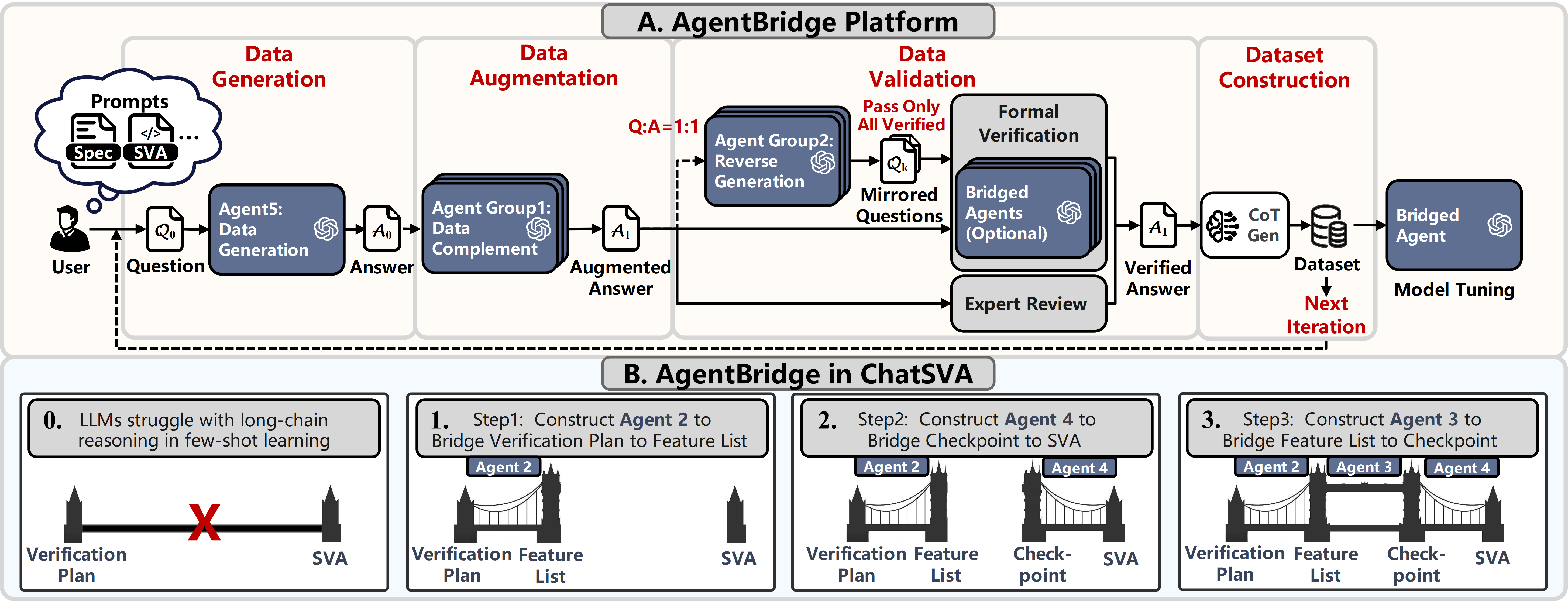} 
  \caption{\small AgentBridge Platform \& AgentBridge in ChatSVA}
  \label{fig:AB}
  \vspace{-4pt}
\end{figure*}
%----------------------------------------------------
%-----------Design
%----------------------------------------------------

\section{ChatSVA}
\label{sec:ChatSVA}
% 简短写法
% As introduced in Section~\ref{sec:Design}, the ChatSVA framework instantiates our long-chain reasoning decomposition strategy. It transforms the monolithic Spec-to-SVA task into a structured, four-stage pipeline that mirrors real-world verification workflows. This pipeline processes data through a series of representations: the initial high-level \textbf{Spec} is converted into a structured \textbf{Verification Plan}; this plan is then decomposed into a \textbf{Feature List} of modular functionalities; each feature is detailed into one or more \textbf{Checkpoints} representing specific test cases; and finally, each checkpoint is translated into executable \textbf{SVA} code.

%Learning from the multi-stage modern hardware verification workflow, where engineers iteratively decompose high-level specifications into verifiable units for precise SVA formulation, we define the following key data representations:

%原写法
As introduced in Section~\ref{sec:Design}, the ChatSVA framework instantiates our long-chain reasoning decomposition strategy. It transforms the monolithic Spec-to-SVA task into a structured, four-stage pipeline, mirroring expert verification workflows (Fig.~\ref{fig:workflow}.A). This decomposition is based on a series of well-defined intermediate data representations that connect each stage. We define these key representations as follows:
(1) Spec: A high-level natural language description of the hardware design. A complete Spec should encompass functional descriptions, input and output ports, internal register mappings, and the temporal behavior of internal signals.  
(2) Verification Plan: A structured intermediate representation that summarizes functional descriptions from the Spec, extracts signal logic relationships and verification requirements, and formats them into a standardized template for downstream processing.
(3) Feature List: The extraction and refinement of design requirements and constraints from the Verification Plan. Feature lists describe the expected functionality of the hardware design under various test scenarios.  
(4) Checkpoint: A detailed, implementation-level description of a feature list. Each checkpoint represents a specific verification point; a single feature list can map to multiple checkpoints. 
(5) SVAs: The code-level implementation of a checkpoint.

\noindent\textbf{ChatSVA Workflow.} 
ChatSVA automates SVA generation through a four-stage LLM pipeline, starting from user-provided Spec. First, the SpecWiz agent (\texttt{Agent1}) extracts verification-critical information to generate structured Verification Plans (Fig.~\ref{fig:workflow}.B.1). This intermediate representation is then processed by the Feature Generator (\texttt{Agent2}) to produce modular Feature Lists (Fig.~\ref{fig:workflow}.B.2). Subsequently, the Checkpoint Generator (\texttt{Agent3}) generates exhaustive, implementation-specific checkpoints from each feature (Fig.~\ref{fig:workflow}.B.3). Finally, the SVA Generator (\texttt{Agent4}) translates individual checkpoints into syntactically correct SVAs, directly embeddable for formal verification or simulation (Fig.~\ref{fig:workflow}.B.4). This collaborative effort by four specialized agents systematically automates the extraction, analysis, and mapping of functional requirements to specific implementations, ensuring high-quality SVA generation.

%----------------------------------------------------
%-----------Design
%----------------------------------------------------

\section{AgentBridge Platform}
\label{sec:AgentBridge}

%Inspired by bridge engineering principles, we introduce the \textbf{AgentBridge} to modularize long-chain reasoning into independently verifiable segments, strategically creating high-quality datasets at these verifiable ``endpoints''. These datasets empower robust ``anchor'' agents, which generate and validate reliable intermediate data, effectively bridging specification and SVA. This approach significantly enhances dataset quality and boosts performance in domain-specific, few-shot contexts.

As introduced in Section~\ref{sec:Design}, AgentBridge addresses the data scarcity challenge in SVA generation. Its design objective is to generate high-quality datasets  across all stages of the ChatSVA pipeline. 
To achieve this, AgentBridge is built upon three foundational principles that guide its data synthesis process.

\subsection{AgentBridge Principles}
\label{sec:form}
The AgentBridge data generation process, which transforms an input set $\mathcal{D}_{\text{in}}$ into an output set $\mathcal{D}_{\text{out}}$, is governed by three principles.

\noindent\textbf{Principle 1: Directional Information Constraint.}
This principle mandates a directional information flow where outputs are functional subsets of the input. This structure ensures that any generated output has a definite ``right-or-wrong'' status against its source, rather than being partially correct. Using a semantic interpretation function, $\text{sem}(\cdot)$, which maps an artifact to its set of functionalities, this is formalized as:
\begin{equation}
\forall a \in \mathcal{D}_{\text{out}}, \exists x \in \mathcal{D}_{\text{in}} \text{ s.t. } a \in \mathcal{G}(x) \land \text{sem}(a) \subseteq \text{sem}(x)
\label{eq:semantic_subset}
\end{equation}
This constraint permits decomposition (e.g., Spec-to-SVA) but precludes synthesis tasks that are logically unsound as they require fabricating information.

\noindent\textbf{Principle 2: Ground Truth Provenance.}
To ensure the integrity of the directional information flow established by Principle 1, the input set $\mathcal{D}_{\text{in}}$ must originate exclusively from a verified ``golden'' dataset, $\mathcal{D}_{\text{gold}}$. A preprocessing function, $\phi$, prepares this golden data for specific generation tasks.
\begin{equation}
\mathcal{D}_{\text{in}} = \phi \left( \mathcal{D}_{\text{gold}} \right) = \{ \phi(d) \mid d \in \mathcal{D}_{\text{gold}} \}
\label{eq:golden}
\end{equation}
This ensures the entire process is anchored to a foundation of error-free data, preventing cascading failures.

\noindent\textbf{Principle 3: Output Verifiability.}
While Principle 1 ensures outputs have a definite status, this principle mandates that a practical method must exist to determine it. We define a universal verification function, $\mathcal{V}(y)$, which encapsulates the hybrid protocol of using either direct methods ($\mathbb{V}_{\text{direct}}$) or a Bridged Agent ($\mathcal{B}$).
\begin{equation}
\mathcal{V}(y) := \mathbb{V}_{\text{direct}}(y) \lor \mathbb{V}_{\text{direct}}(\mathcal{B}(y))
\label{eq:verifiability}
\end{equation}
The outcome of $\mathcal{V}(y)$ is boolean (True/False), guaranteeing that the status of an output can be practically determined.

\subsection{Workflow of AgentBridge Platform} \label{sec:AgentBridge Platform} The AgentBridge workflow, depicted in Fig.~\ref{fig:AB}.A, is a self-improving closed loop that systematically generates high-purity datasets. The process begins with \textbf{Data Generation}, where an input $Q_0$ from golden data (Principle 2) is processed by \texttt{Agent5} to produce answer candidates $A_0 = \{a_1, ..., a_N\}$. This $1:N$ decomposition adheres to Principle 1, ensuring each candidate $a_i$ is a verifiable functional subset of $Q_0$. Next, in the \textbf{Data Augmentation} stage, \texttt{Agent Group1} analyzes the $Q_0 \to A_0$ mapping to identify and fill coverage gaps, merging its outputs into an augmented set $A_1$. An "Agent Group" denotes a process where each element of a set is independently processed. The subsequent \textbf{Data Validation} stage employs a hybrid protocol (Principle 3). For general $1:N$ tasks, each item in $A_1$ is directly verified. For $1:1$ mappings, the workflow uses a robust reverse-generation check: \texttt{Agent Group2} produces mirrored questions $Q_k$ from $A_1$, and the set $A_1$ is accepted only if all questions in $Q_k$ pass verification. Validation methods are task-specific, including formal verification, expert review, or transformation via Bridged Agents. Finally, in \textbf{Dataset Construction}, validated candidates are paired with the original question $Q_0$. A Chain-of-Thought (CoT) prompt is generated by documenting the successful validation path. This CoT and the final QA pairs form the dataset used to fine-tune all internal agents, including Bridged Agents, thus closing the self-improving loop.

\begin{figure*}[h!] % 使用h!强制当前位置

  \centering
  \includegraphics[width=0.93\linewidth]{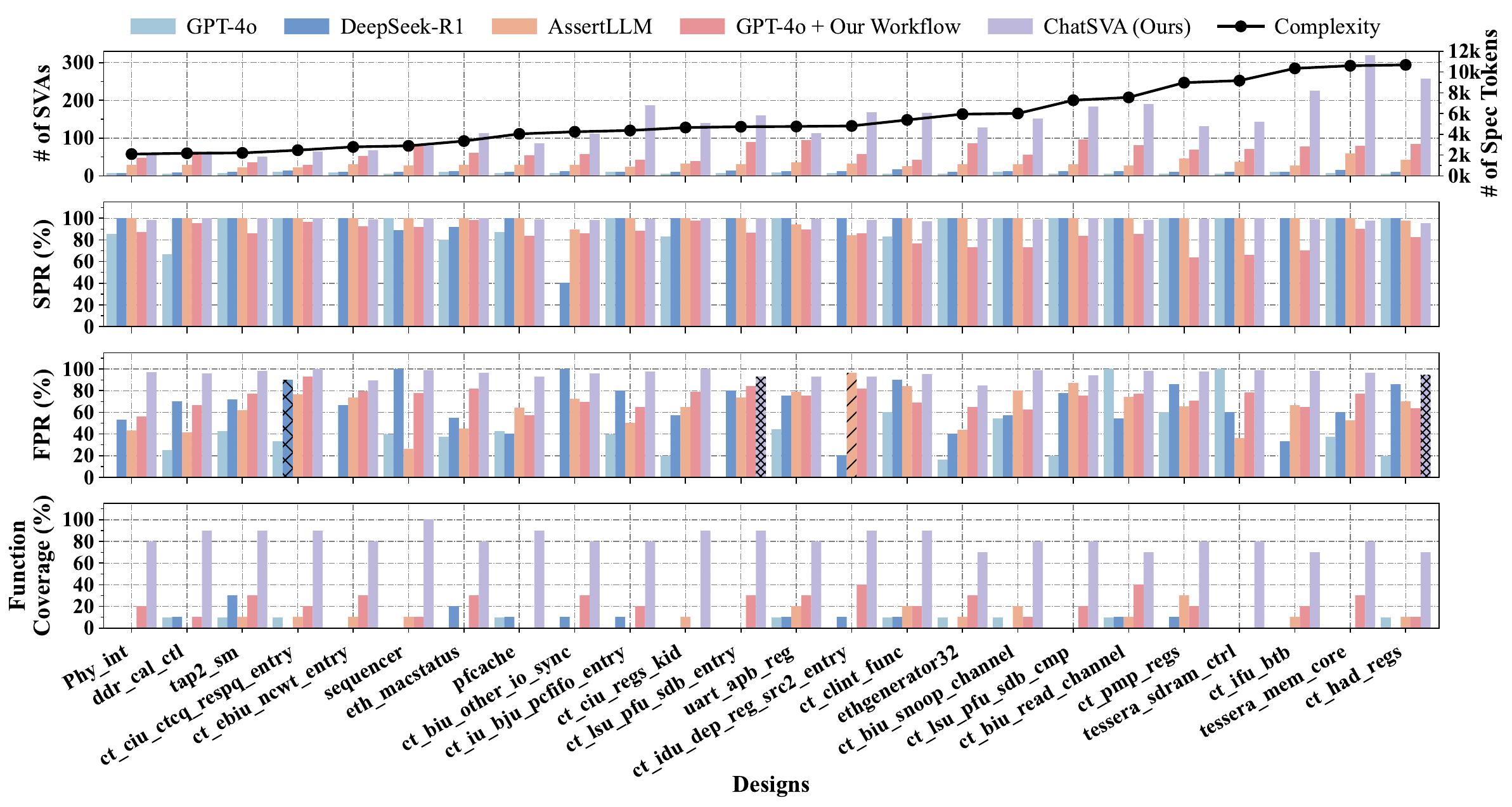} % 微调宽度
  \caption{\small ChatSVA Performance Comparisons}
  \label{fig:performence}
  \vspace{-5pt}
\end{figure*}

\subsection{Instantiating ChatSVA with AgentBridge}
\label{sec:AgentBridge in ChatSVA}

AgentBridge is instantiated to construct datasets for the three specialized agents in the ChatSVA framework, as depicted in Fig.~\ref{fig:AB}.B.
The process begins with \texttt{Agent2}. To satisfy Principle 2, we establish a golden input by structuring Spec into expert-verified Verification Plans. \texttt{Agent2} then decomposes these plans into Feature Lists. The outputs are subsequently verified via expert review, a direct verification method under Principle 3.
Next, we strategically prioritize the dataset for \texttt{Agent4} due to its critical $1:1$ mapping. This bijective task, a special case of Principle 1, enables a robust reverse-generation validation: a Checkpoint is generated from a golden SVA to form a candidate pair. To validate this pair, the generated Checkpoint is used to reverse-generate a new SVA. The original pair is confirmed as correct only if this new SVA is functionally equivalent to the original golden SVA, a process reinforced by formal verification.
Finally, the fine-tuned \texttt{Agent4} serves as a Bridged Agent to construct the dataset for \texttt{Agent3}, resolving the challenge of data validation. For each Checkpoint generated by \texttt{Agent3}, \texttt{Agent4} transforms it into a verifiable SVA. This closed-loop validation strategy makes the Checkpoint generation task concretely verifiable. 

\section{Evaluations}
\label{sec:eval}

\subsection{Experimental Setup}
\label{sec:setup}

\subsubsection{\textbf{Model Training}}
% \label{sec:proto}
% \paragraph{\textbf{Model Training.}}

% \noindent\textbf{Model Training.}

We constructed a 15.36 GB dataset for SFT and RAG, comprising data from all pipeline stages enriched with domain knowledge and verification guidelines. The Llama3.1-8B models for \texttt{Agent2-4} underwent full-parameter SFT using LlamaFactory~\cite{llamafactory} on an 8$\times$A800 GPU server. Key hyperparameters included a sequence length of 8192, a learning rate of 1e-5, and a batch size of 64 for 3 epochs. For high-level reasoning, \texttt{Agent1} utilizes GPT-4o with RAG. All agents were configured with a temperature of 0.2.

\subsubsection{\textbf{Baselines and Benchmark.}}
% \label{sec:benchmark}

% We evaluated ChatSVA on 24 golden RTL designs from the hardware verification benchmark named FIXME~\cite{FIXME}, which includes an SVA generation task independent of the training set. These designs represent a diverse range of industrial digital circuits, such as control units, data paths, and state machines, with RTL code lengths varying from 100 to 800 lines.

% \paragraph{\textbf{Baselines and Benchmark.}}

% \noindent\textbf{Baselines and Benchmark.}
We evaluate ChatSVA against general-purpose models (GPT-4o, DeepSeek-R1) and AssertLLM~\cite{assertllm}. AssertLLM is selected as the primary SOTA baseline as it is also a multi-agent framework that generates SVAs from Spec, making it a direct counterpart to our approach. As established in Section~\ref{sec:background}, this comparison intentionally excludes methods that require RTL code as input to avoid potential co-source errors. To ensure a fair comparison, we reproduced the AssertLLM methodology using GPT-4o as its base model. All methods are evaluated on the \textbf{FIXME} benchmark~\cite{FIXME}, a comprehensive public test set featuring diverse industrial designs (e.g., CPUs, GPUs, IPs).

  \begin{figure*}[t]
  
\centering
\includegraphics[width=0.90\linewidth]{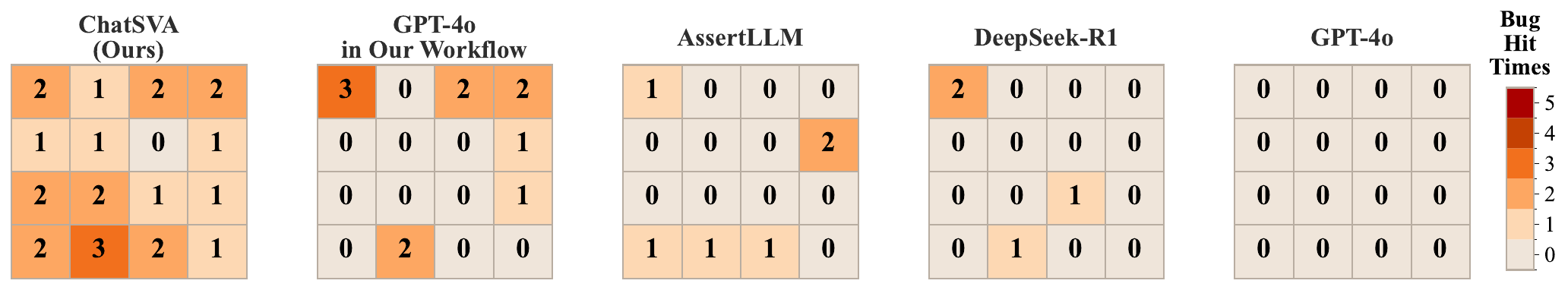}
  \caption{\small Distribution of Bug Detection}
  \label{fig:Bug_Distribution}
  \vspace{-5pt}
\end{figure*}

\begin{figure}[t] % 浮动体定位控制符

  \centering
  \includegraphics[width=0.90\columnwidth]{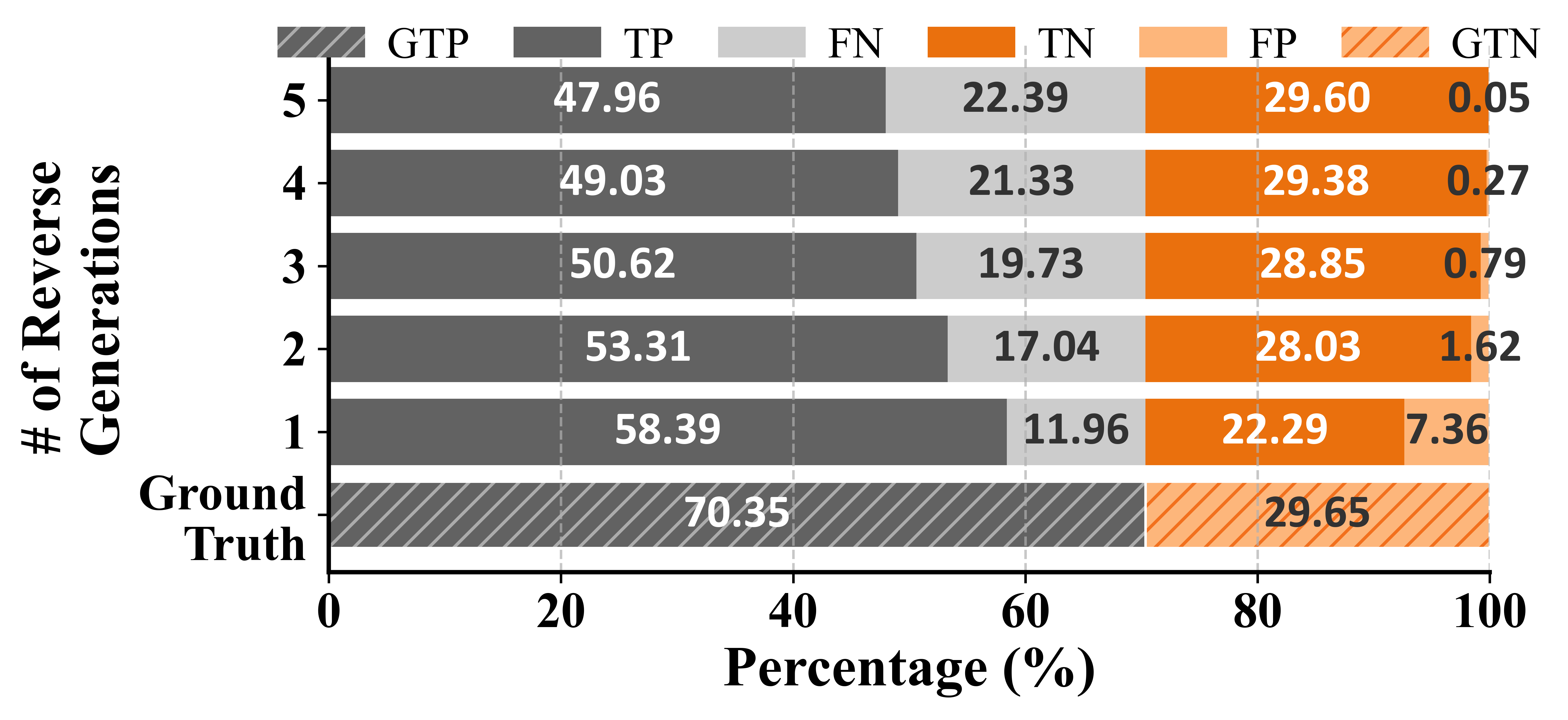} % 宽度适配单栏
  \caption[Short caption]{\small Data Distribution in Reverse Generation Method}
  \label{fig:stackbar}
  \vspace{-5pt}
\end{figure} 

 \subsubsection{\textbf{Evaluation Metrics}}
% \label{sec:metrics}

% \paragraph{\textbf{Baselines and Benchmark.}}
% \noindent\textbf{Evaluation Metrics.}
% We define three key metrics for SVA quality: \textbf{Syntax Pass Rate (SPR)}, \textbf{Function Pass Rate (FPR)}, and \textbf{Function Coverage}. 
% These are calculated using the general Pass Rate formula (Eq.~\ref{eq:general_metric}), which leverages the verification function $\mathcal{V}$ from Eq.~\ref{eq:verifiability}. For calculation, the boolean outcome of $\mathcal{V}$ is treated as 1 or 0.
% These use the general Pass Rate formula (Eq.~\ref{eq:general_metric}), treating the boolean verification function $\mathcal{V}$ (Eq.~\ref{eq:verifiability}) as 1 or 0.
%  SPR measures the proportion of syntactically correct SVAs. FPR assesses the proportion of functionally correct SVAs among those that are syntactically valid. Function Coverage evaluates the bug-detection capability of functionally correct SVAs against predefined bug types. (Protocol Violations, Illegal Branch, etc.)
% \begin{equation}
% \text{PR} = \frac{\sum_{i=1}^{N} \mathcal{V}(n_i)}{N} \times 100\%
% \label{eq:general_metric}
% \end{equation}

We define three key metrics for SVA quality: \textbf{Syntax Pass Rate (SPR)}, \textbf{Function Pass Rate (FPR)}, and \textbf{Function Coverage}. These are calculated using the general Pass Rate formula (Eq.~\ref{eq:general_metric}), which leverages the verification function $\mathcal{V}$ from Eq.~\ref{eq:verifiability}. For calculation, the boolean outcome of $\mathcal{V}$ is treated as 1 or 0. SPR measures the proportion of syntactically correct SVAs. FPR assesses the proportion of functionally correct SVAs among those that are syntactically valid. Function Coverage evaluates the bug-detection capability of functionally correct SVAs against predefined bug types. (Protocol Violations, Illegal Branch, etc.)
\begin{equation}
\text{PR} = \frac{\sum_{i=1}^{N} \mathcal{V}(n_i)}{N} \times 100\%
\label{eq:general_metric}
\end{equation}

To validate the capability of AgentBridge in generating high-purity training data, we measure its Data Precision ($\mathcal{P}$). 
We begin with a set of checkpoint candidates generated from over 500 golden SVAs, which are manually labeled to establish ground truth—either \emph{Ground Truth Positive} (GTP) or \emph{Ground Truth Negative} (GTN). 
AgentBridge then validates these same candidates, classifying them as \emph{Validation Positive} (VP) or \emph{Validation Negative} (VN). This process yields four outcomes: True Positives (TP) for correctly validated GTP items, False Positives (FP) for incorrectly validated GTN items, and their counterparts, True Negatives (TN) and False Negatives (FN). Since FPs represent harmful, hallucinated data that could contaminate the training set, precision $\mathcal{P} \mathcal{=} \frac{\text{TP}}{\text{TP} + \text{FP}}$, is the critical metric for evaluating the filtering effectiveness of the platform.

\subsection{Results and Analysis}
\label{sec:result}

\subsubsection{\textbf{SVA Generation Quality and Efficacy.}}
% \noindent\textbf{SVA Generation Quality and Efficacy.}
% As shown in Fig.~\ref{fig:performence}, ChatSVA establishes a new SOTA in SVA generation, producing an average of 139.50 SVAs per design with a \textbf{98.66\%} SPR, a \textbf{96.12\%} FPR, and \textbf{82.50\%} Function Coverage. This performance surpasses all baselines.
% Compared to general-purpose LLMs, ChatSVA achieves a 19.80$\times$ and 14.14$\times$ improvement over GPT-4o (4.17\%) and DeepSeek-R1 (5.83\%) in Function Coverage, with SVA generation volume also increasing by over 18$\times$ against GPT-4o. This demonstrates that without a guiding methodology, even powerful general-purpose LLMs fail to generate comprehensive and functionally correct SVAs.
% More importantly, ChatSVA significantly outperforms AssertLLM. While AssertLLM has a high SPR (98.52\%), its FPR and Function Coverage are only 62.84\% and a mere 7.50\%, respectively. Consequently, ChatSVA achieves an \textbf{11$\times$} improvement in Function Coverage over AssertLLM, a substantial leap in functional verification. The low functional performance of AssertLLM, comparable to general-purpose models, indicates its multi-agent structure, while a step in the right direction, lacks mechanisms for effective reasoning decomposition and data provisioning. ChatSVA overcomes these limitations with our S1 and S2 strategies.

As shown in Fig.~\ref{fig:performence}, ChatSVA establishes a new SOTA in SVA generation, producing an average of 139.50 SVAs per design with a \textbf{98.66\%} SPR, a \textbf{96.12\%} FPR, and \textbf{82.50\%} Function Coverage. This performance surpasses all baselines.
Compared to general-purpose LLMs, ChatSVA achieves a 19.80$\times$ and 14.14$\times$ improvement over GPT-4o (4.17\%) and DeepSeek-R1 (5.83\%) in Function Coverage, with SVA generation volume also increasing by over 18$\times$ against GPT-4o. This demonstrates that without a guiding methodology, even powerful general-purpose LLMs fail to generate comprehensive and functionally correct SVAs.
More importantly, ChatSVA significantly outperforms AssertLLM. While AssertLLM has a high SPR (98.52\%), its FPR and Function Coverage are only 62.84\% and a mere 7.50\%, respectively. Consequently, ChatSVA achieves an \textbf{11$\times$} improvement in Function Coverage over AssertLLM, a substantial leap in functional verification. The low functional performance of AssertLLM, comparable to general-purpose models, indicates its multi-agent structure, while a step in the right direction, lacks mechanisms for effective reasoning decomposition and data provisioning. ChatSVA overcomes these limitations with our S1 and S2 strategies.

\subsubsection{\textbf{Ablation Study.}}
% \noindent\textbf{Ablation Study.}
% To deconstruct the sources of this performance gain, we conducted ablation studies isolating the contributions of our two core strategies: long-chain reasoning decomposition (S1) and domain-specific data synthesis (S2).

% First, to validate the power of S1, we applied our workflow to GPT-4o without any fine-tuned models. Under the workflow guidance, the base GPT-4o improves its FPR from 43.09\% to 72.88\% (+29.79pp) and boosts Function Coverage by nearly 5$\times$, increasing from 4.17\% to 20.83\%. In addition, the workflow-guided GPT-4o surpasses AssertLLM on all functional metrics, achieving +10.04pp higher FPR and +13.33pp higher Function Coverage, even though both methods rely on the same GPT-4o model. These results indicate that the ChatSVA workflow itself is a primary driver of performance: by explicitly structuring the long-chain reasoning process, it delivers stronger functional gains than prior frameworks.

% Next, to demonstrate the criticality of S2, we compare the GPT-4o guided by the complete ChatSVA workflow. With AgentBridge data synthesis and subsequent fine-tuning, ChatSVA further raises Function Coverage from 20.83\% to 82.50\%, which is nearly a 4$\times$ improvement over the workflow-guided GPT-4o. This shows that while S1 provides a strong foundation for reasoning and decomposition, S2 is indispensable for learning hardware-verification semantics at scale and for reaching SOTA performance.

To deconstruct the sources of this performance gain, we conducted ablation studies isolating the contributions of our two core strategies: long-chain reasoning decomposition (S1) and domain-specific data synthesis (S2).
First, to validate the power of S1, we applied our workflow to GPT-4o without any fine-tuned models. Our workflow boosts the FPR of the base GPT-4o from 43.09\% to 72.88\% (+29.79pp) and improves Function Coverage by nearly 5$\times$ (from 4.17\% to 20.83\%). Furthermore, this workflow-guided model outperforms AssertLLM in all functional metrics (FPR: +10.04pp; Function Coverage: +13.33pp), despite both using the same GPT-4o base model. This proves that the ChatSVA workflow  by structuring the long-chain reasoning process, is a primary driver of performance and is superior to prior frameworks.
Next, to demonstrate the criticality of S2, we compare the GPT-4o in our workflow with the full ChatSVA framework. The full ChatSVA framework further improves Function Coverage from 20.83\% to 82.50\%—a nearly 4$\times$ increase over the workflow-guided GPT-4o. This highlights that while a superior workflow (S1) provides a strong foundation, fine-tuning on the high-purity data generated by AgentBridge (S2) is indispensable for mastering the complex semantics of hardware verification and achieving SOTA performance.

\subsubsection{\textbf{Qualitative Analysis}}

% \label{sec:bug_distrubution}
% \noindent\textbf{Qualitative Analysis.}
% To further assess the practical quality of bug detection beyond average scores, we analyzed the distribution of bug detection in Fig.~\ref{fig:Bug_Distribution}. ChatSVA demonstrates systematic and comprehensive coverage, successfully identifying 15 out of 16 unique bug types. The even distribution of these detections indicates that ChatSVA generates a diverse set of SVAs targeting a wide range of functional logic, making it a reliable verification partner.In stark contrast, all baseline models fail to provide dependable coverage. General-purpose models such as GPT-4o and DeepSeek-R1 are largely ineffective, while AssertLLM exhibits only sporadic success by identifying just 5 of 16 bug types. This unreliable performance confirms that only the complete ChatSVA framework provides the broad and balanced coverage essential for practical hardware verification.

To assess the practical bug-detection quality beyond average scores, we analyze the distribution of detected bugs in Fig.~\ref{fig:Bug_Distribution}. ChatSVA demonstrates systematic and comprehensive coverage, successfully identifying 15 out of 16 unique bug types. The even distribution of these detections indicates that ChatSVA generates a diverse set of SVAs targeting a wide range of functional logic, making it a reliable verification partner.In stark contrast, all baseline models fail to provide dependable coverage. General-purpose models like GPT-4o and DeepSeek-R1 are largely ineffective, while AssertLLM exhibits only sporadic success by identifying just 6 of 16 bug types. This unreliable performance confirms that only the complete ChatSVA framework provides the broad and balanced coverage essential for practical hardware verification.

%To further assess the practical quality of bug detection beyond average scores, we analyzed the distribution of bug detection in Fig~\ref{fig:Bug_Distribution}. ChatSVA demonstrates systematic and comprehensive coverage, successfully identifying 15 of 16 unique bug types. The even distribution of detections indicates that ChatSVA generates a diverse set of SVAs targeting a wide range of functional logic, making it a reliable verification partner. In stark contrast, models without domain-specific fine-tuning exhibit inconsistent performance. For instance, DeepSeek-R1 in ChatSVA Workflow over-concentrates on specific bugs while completely failing to detect many others. Baseline models are largely ineffective. This analysis confirms that ChatSVA provides the balanced coverage essential for practical verification.

\subsubsection{\textbf{Validation of AgentBridge}}
High-purity training data is essential for our SFT approach, and we validate efficacy of AgentBridge in producing such data. AgentBridge leverages a novel reverse generation mechanism (\texttt{Agent Group2}) using $k$ independent agents to filter hallucinatory False Positives (FPs) that could poison the dataset. As Fig.~\ref{fig:stackbar} shows, increasing the number of agents from $k=1$ (yielding a significant 7.36\% FP rate) to $k=5$ (nearly eliminating FPs to just 0.05\%) boosts Data Precision ($\mathcal{P}$) from 88.8\% to a near-perfect 99.9\%. This represents a deliberate trade-off prioritizing purity over completeness.

%%%%%%%%%%%%%%%%%%%%%这里是rn node+cva6的后端信息，根据需要调整即可%%%%%%%%%%%%%%%%%%%%

% \begin{figure}[!t]
%   \centering
%   \includegraphics[height=1.2in, width=2.9in]{5.Evaluation_figure/noc_bandwidth.png}
%   \caption{NoC Bandwidth}
%   \label{fig:bw_link}
% \end{figure}

%\begin{table}[!t]
%\footnotesize
%\centering
%    \begin{minipage}{0.49\textwidth}
%	\caption{table}
%	\label{tab:ppa}
%	\begin{tabular}{|p{22mm}|c|c|}
%		\hline
%		\textbf{Design}   & \textbf{Comupte Unit} & \textbf{Coherence Unit}  \\
%	    \hline \hline
%		PDK & \multicolumn{2}{|c|}{Skywater 130 nm CMOS}  \\
%		\hline
%		PVT & \multicolumn{2}{|c|}{sky130\_fd\_sc\_hd\_tt\_025C\_1v80}\\
%		\hline
%		Frequency & \multicolumn{2}{|c|}{100MHz} \\
%  		\hline
%		Die Area & $10.28mm^{2}$ & $7.21mm^{2}$ \\
%              \hline
%		Macro & 1 (private \$ + router ) & 18 (SRAM)\\
%            \hline
%		Total Macro Area & $4.46mm^{2}$ & $1.53mm^{2}$ \\
%            \hline
%		Cell Count & 460K & 270K\\
 %           \hline
%            Dynamic Power & 235.30mW & 105.54mW\\
%		\hline
%	\end{tabular}
%	\end{minipage}
%\end{table}

% \begin{figure}[t]
%   \centering
%   \includegraphics[width=3in]{5.Evaluation_figure/50Layout_2.png}
%   \caption{Physical Layout With Commercial Configuration}
%   \label{fig:commercial_layout}
% \end{figure}

%=====for journal extend==========
% \subsection{Analysis}

% % \subsubsection{Improvements needed for current open EDA tools}

% \subsection{Future opportunities}

\section{Conclusion}
\label {sec:con}
 % Automating SVA generation for hardware verification is challenged by general LLMs' functional accuracy and data scarcity. We propose ChatSVA, an end-to-end multi-agent system, leveraging AgentBridge for high-purity data generation and validation in few-shot, long-chain reasoning tasks. Experiments demonstrate ChatSVA achieves SOTA performance with 98.66\% SPR, 96.12\% FPR and 82.50\% Function Coverage, yielding over 14$\times$ improvement in bug detection over baselines.  
 % This work advances agile verification and positions AgentBridge as a generalizable dataset construction method in IC verification domains.

%Ver.2
In this work, we propose ChatSVA, an end-to-end multi-agent system, leveraging AgentBridge for high-purity data generation and validation in few-shot, long-chain reasoning tasks. Experiments demonstrate that ChatSVA achieves SOTA performance with 98.66\% SPR, 96.12\% FPR and 82.50\% Function Coverage, yielding over 11$\times$ improvement in bug detection over the previous SOTA.
This work advances agile verification and positions AgentBridge as a generalizable dataset construction method in IC verification domains. 

\begin{acks}
This work is supported by the National Natural Science Foundation of China (NSFC Grant No.\ 92464301), the National Key Research and Development Program (Grant No.\ 2024YFB4405600), and the Key Research and Development Program of Jiangsu Province (Grant No.\ BG2024010).
\end{acks}

\bibliographystyle{ACM-Reference-Format}
\bibliography{reference}

\end{document}